\documentclass{cmspaper}
\begin{document}

%==============================================================================
% title page for few authors

\begin{titlepage}

% select one of the following and type in the proper number:
  \conferencereport{2008/112}
   \date{16 December 2008}

  \title{Forward physics with CMS}

  \begin{Authlist}
      Monika Grothe\Aref{a}
      \Instfoot{UW}{University of Wisconsin, Madison, USA}
  \end{Authlist}

% if needed, use the following:
%\collaboration{Flying Saucers Investigation Group}
%\collaboration{CMS collaboration}

  \Anotfoot{a}{monika.grothe@cern.ch}

  \begin{abstract}
Forward physics with CMS at the LHC covers a wide range of physics subjects, including 
very low-$x_{Bj}$ QCD, underlying event and multiple interactions characteristics, 
$\gamma$-mediated processes, shower development at the energy scale of primary cosmic ray 
interactions with the atmosphere, diffraction in the presence of a hard scale and even 
MSSM Higgs discovery in central exclusive production. Selected 
feasibility studies to illustrate the forward physics potential of CMS are presented.

  \end{abstract} 

% if needed, use the following:
\conference{Presented at {\it 2008 Physics at LHC}, Split, Croatia, Sept 29 - Oct 4, 2008}
\submitted{Submitted to {\it Proceedings of Science}}
%\note{Preliminary version}
  
\end{titlepage}

\setcounter{page}{2}%JPP

%==============================================================================

\section{Introduction}

Forward physics at the LHC covers a wide range of diverse physics subjects that have in 
common that particles produced at small polar angles, $\theta$, and hence large 
values of rapidity provide a defining characteristics. At the Large-Hadron-Collider (LHC), 
where proton-proton collisions occur at 
center-of-mass energies of 14 TeV, the maximal possible rapidity is 
$y_{max} = \ln{\frac{\sqrt{s}}{m_{\pi}}}\sim 11.5$. The central components of CMS are 
optimized for efficient detection of processes with large
polar angles and hence high transverse momentum, $p_T$.
They extend down to about $|\theta| = 1^\circ$ from the beam axis
or $|\eta| = 5$, where $\eta = - \ln{[\tan{( \theta / 2 )} ] }$ is the pseudorapidity.
In the forward region, the central CMS components are complemented by several CMS 
and TOTEM subdetectors
with coverage beyond $|\eta| =5$. TOTEM is an approved experiment at 
the LHC for precision measurements of the $pp$ elastic and total cross sections.
The combined CMS and TOTEM apparatus comprises two suites of calorimeters with tracking 
detectors in front plus near-beam proton taggers. The CMS Hadron Forward (HF) calorimeter 
with the TOTEM telescope T1 in front 
covers the region $3 < |\eta | < 5$, the CMS CASTOR calorimeter with the TOTEM telescope 
T2 in front covers $5.2 < |\eta| < 6.6$. The CMS ZDC calorimeters are
installed at the end of the straight LHC beam-line section, at a distance of 
$\pm 140$~m from the interaction point (IP). 
Near-beam proton taggers will be installed by TOTEM at 
$\pm 147$~m and $\pm 220$~m from the IP.
The kinematic coverage of the combined CMS and TOTEM apparatus is unprecedented at a
hadron collider. The CMS and TOTEM collaborations have described the considerable physics 
potential of joint data taking in a report to the LHCC \cite{opus}. 
Further near-beam proton taggers in combination with very fast timing detectors to be
installed at $\pm 420$~m from the IP (FP420) are in the proposal stage in CMS.
FP420 would give access to possible discovery processes in forward physics at the 
LHC~\cite{fp420}.

\section{Physics with forward detectors}

The CMS HF, CASTOR and ZDC 
calorimeters~\cite{PTDR1} 
and the TOTEM T1 and T2 telescopes~\cite{TOTEM} are particulatly suited for 
QCD studies at very low values of Bjorken-$x$, which have the potential of differentiating
between DGLAP-type and BFKL-like parton evolution and of identifying effects of parton
saturation.
Measurements at the
HERA $ep$ collider have explored low-$x_{Bj}$ dynamics down to values of a few $10^{-5}$.
At the LHC the minimum accessible $x$ decreases by a factor $\sim 10$ for each
2 units of rapidity. A process with a hard scale of $Q \sim 10$~GeV and within the 
acceptance of T2/CASTOR ($\eta = 6$) can occur at $x$ values as low as 
$10^{-6}$.

An example is Drell-Yan production of $e^+ e^-$ pairs, 
$ q q \rightarrow \gamma^\star \rightarrow e^+ e^-$, which probes primarily 
the quark content of the proton. 
At values sufficiently low in $x$ non-linear QCD effects of parton recombination and 
saturation within the proton are expected to set in, taming the 
rapid rise of the proton gluon density observed by HERA and preserving unitarity.
Generator-level studies~\cite{opus} indicate that saturation effects may 
manifest themselves as production cross sections for Drell-Yan electrons within the 
T2/CASTOR acceptance that are sizably different from those
predicted in models without saturation.

Another example of a low-$x$ process is production of forward jets, observable in HF or 
CASTOR. The CMS HF calorimeter offers excellent jet reconstruction capabilities. Forward
jets in HF reach a $p_T$ resolution of $\sim 19\% $ at 20~GeV/c and $\sim 10\% $ at 100 GeV/c 
\cite{PASfwdjets}. Particularly sensitive to
BFKL-like QCD evolution dynamics are dijets with large rapidity separation, which 
enhances the available phase space for BFKL-like parton radiation between the jets. 
Generator-level studies for these so-called Muller-Navelet-type dijets with 
$\Delta \eta$ up to 9 are discussed in~\cite{Salim}.
Likewise dijets separated by a large rapidity gap are of interest since they indicate
a process in which no color flow occurs in the hard scatter but where, contrary to the 
traditional picture of soft Pomeron exchange, also a high transverse momentum transfer 
occurs across the gap. 

Another area where the CMS forward calorimeters and TOTEM telescopes are essential tools
is Monte Carlo tuning.
The hard scatter in hadron-hadron collisions takes place in a dynamic environment,
refered to as the ``underlying event'' (UE), where
additional soft or hard interactions between the partons and 
initial and final state radiation occur. The effect of the UE can only be described 
by means of tuning Monte Carlo multiplicities 
and energy flow predictions to data. 
As shown in~\cite{Borras}, the forward detectors are sensitive
to features of the UE that central detector information alone cannot constrain.
High uncertainties are also present in modelling the interaction of primary cosmic rays 
of PeV energy with the atmosphere. Their rate of occurance per year is too low for
reliable quantitative analysis. The center-of-mass energy in $pp$ collisions at the LHC 
corresponds to 100 PeV energy in a fixed target collision. As discussed in~\cite{opus},
measurements of energy and particle flow with T2/CASTOR and ZDC will be able to constrain 
significantly the Monte Carlo models currently used in the cosmic rays community.

\section{Physics with a veto on the forward detetcors}

Events of the type $pp \rightarrow pXp$ or $pp \rightarrow Xp$, where no color exchange
takes place between the proton(s) and the system $X$, can be caused by $\gamma$ exchange,
or by diffractive interactions. In both cases, the absence of color flow between the
proton(s) and the system $X$ results in a large gap in the rapidity distribution of the
hadronic final state. In the following, we discuss, for several exemplary processes, how 
the forward detectors can be used to detect such hadronic final states.

Diffraction can occur with a hard scale. In that case
perturbative QCD 
(pQCD) allows the cross sections for these processes
to be factorized into that one of the hard scatter and a diffractive particle 
distribution function (dPDF). In diffractive hadron-hadron scattering, rescattering between
spectator particles breaks the factorization. The so-called rapidity gap survival 
probability, $< | S^2 | >$, quantifies this effect~\cite{survival} and can be measured by means of the ratio
of diffractive to inclusive processes with the same hard scale. At the Tevatron, the ratio 
is found to be ${\cal O}(1 \%)$~\cite{tevatron}.
Theoretical expectations for $< | S^2 | >$ the LHC vary from a fraction of a percent to as much as 
${\cal O}(25\%)$~\cite{predLHC}. 

In hard single diffraction (SD) at LHC energies, the complication arises that
the rapidity gap is generally boosted to large values of rapidity. A gap within the 
acceptance of the HF and CASTOR calorimeters limits the values of $\xi$, the 
fractional momentum loss of the diffractively scattered proton, to $\xi < 0.01$. 
Two processes have been studied in detail~\cite{SDW, SDjj},
SD dijet production, $pp \rightarrow pjjX$, sensitive to the dPDF gluon content, and 
SD $W$ production, $pp \rightarrow pWX$ ($W \rightarrow \mu \nu$), 
sensitive to the dPDF quark content. Both studies assume single-interaction data, i.e. 
absence of event pile-up. The event selection is based on the two dimensional 
distributions of towers with above-noise
activity on the gap side, a. in HF versus CASTOR, or b. in a high-$\eta$ slice versus a 
low-$\eta$ slice of HF towers. The gap side is defined as the side with the lower hadronic 
activity, a definition that can be equally applied to non-diffractive events.

After event selection, the two-dimensional multiplicity distributions on the selected 
gap side show clearly visible differences between the Pyhtia-only prediction (no hard SD)
and the Pythia plus Pomwig (hard SD) one. Assuming  $< | S^2 | > \approx 5\% $ in 
Pomwig, a particularly clear excess is present in the [0,0] bins. For SD $W$ production,
${\cal O}(100)$ events per 100~$\rm pb^{-1}$ are expected in [n(CASTOR), n(HF)] = [0,0], 
with a ratio of SD to non-diffractive events of $\sim 20$. Even more favorable is 
the situation for SD dijet production, where ${\cal O}(300)$ events per 10~$\rm pb^{-1}$ 
are expected, 
with a ratio of $\sim 30$. Dissociation of the 
diffractively scattered proton would lead to a signal enhancement in the [0,0] bin
of $\sim 30\%$ for both channels. The studies assume that CASTOR will be available 
only on one side in the first phase of CMS data taking. A second CASTOR in the opposite 
hemisphere and the use of T1, T2 will improve the observable excess further. 
A method to establish that the observed population of the [0,0] bins is indeed indicative
of the presence of SD events in the data is described in~\cite{SDjj}. The method is
based on the observation that the size of the SD signal in the [0,0] bins can be 
controlled in a predictable way when the cuts for enhancing the SD signal are modified.
Observation of signals in the [0,0] bins of the size given above would already exclude 
values of  $< | S^2 | >$ at the low end of the spectrum of 
theoretical predictions. 

Exclusive dimuon and dielectron production with no significant additional
activity in the CMS detector occurs with high cross section in
$\gamma$-mediated processes at the LHC, either as the pure QED process 
$\gamma \gamma \rightarrow ll$ 
or in $\Upsilon$ photoproduction. The event selection in both cases~\cite{ExclDileptons} 
is based on requiring 
that outside of the two leptons, no other 
significant activity is visible within the central CMS detector, neither in the calorimeter
nor in the tracking system. 

For the pure QED process, in 100~$\rm pb^{-1}$ of single interaction data, ${\cal O} (700)$
events in the dimuon channels and ${\cal O} (70)$ in the dielectron channel can be selected.
Events in which one of the  protons dissociates 
are the dominant source of background and are comparable in statistics to the signal. 
This background can be significantly reduced by means of a veto on activity in CASTOR
and ZDC, by 2/3 in a configuration with a ZDC on each side and a CASTOR on only one side of the IP.
The theoretically very precisely known cross section of this (almost) pure QED process 
is an ideal calibration channel. With
100~$\rm pb^{-1}$ of single-interaction data, an absolute luminosity calibration with ${\cal O}(5\%)$ 
precision is feasible.
Futhermore, exclusive dimuon production is an ideal alignment channel with high statistics 
for the proposed proton taggers at 420~m from the IP. 

Upsilon photoproduction can constrain 
QCD models of diffraction.
Assuming the STARLIGHT 
Monte Carlo cross section prediction, the 1S, 2S and 3S resonances
will be clearly visible in 100~$\rm pb^{-1}$ of single interaction 
data~\cite{ExclDileptons}. 
By means of $p_T^2 (\Upsilon)$ as estimator of the transfered four-momentum 
squared, $t$, at the proton vertex, it might be possible to measure the $t$ dependence of the 
cross section. This dependence is sensitive to the two-dimensional gluon distribution of the 
proton and would give access to the generalized parton distribution function (GPD) of the 
proton.

\section{Physics with near-beam proton taggers}
 
The LHC beamline with its magnets is essentially a
spectrometer in which protons slightly off the beam momentum are bent sufficiently to be 
detectable by means of detectors inserted into the beam-pipe. 
At high luminosity at the LHC, proton tagging is the only means of detecting diffractive and 
$\gamma$ mediated processes because areas of low or no hadronic activity in the detector are
filled in by particles from overlaid pile-up events. 

The TOTEM proton taggers at $\pm 220$~m at nominal LHC optics have acceptance
for scattered protons from the IP for $0.02 < \xi < 0.2$.
Smaller values of $\xi$, $0.002 < \xi < 0.02$, can be achieved with proton taggers at 
$\pm 420$~m. The FP420 proposal~\cite{fp420} foresees employing 3-D Silicon, an
extremely radiation hard novel Silicon technology, for the proton taggers, and additional 
fast timing Cherenkov detectors for the rejection of protons from pile-up events. The 
proposal is currently under consideration in CMS. If approved, installation could
proceed in 2010, after the LHC start-up.

Forward proton tagging capabilities enhance the physics potential of CMS. They would
render possible a precise measurement of the mass and quantum numbers of the Higgs boson
should it be discovered by traditional searches. They also augment the CMS discovery reach
for Higgs production in the minimal supersymmetric extension (MSSM) of the Standard Model 
(SM) and for physics beyond the SM in $\gamma p$ and $\gamma \gamma$ interactions.
A case in point is the central exclusive production (CEP) process~\cite{CEP}, 
$pp \rightarrow p + \phi + p$, where the plus sign denotes the absence of hadronic 
activity between the outgoing protons, which survive the interaction intact, and the 
state $\phi$. The final state consists solely of the
scattered protons, which may be detected in the forward proton taggers, and the decay 
products of $\phi$ which can be detected in the central CMS detector. 
Selection rules force the produced state $\phi$ to have $J^{CP} = n^{++}$ with $n =0, 2, ..$.This process offers hence an experimentally very clean 
laboratory for the discovery of any particle with these quantum numbers that couples 
strongly to gluons. Additional advantages are the possibility to determine the mass of the state
$\phi$ with excellent resolution from the scattered protons alone, independent of its
decay products, and the possibility, unique at the LHC, to determine the quantum numbers of 
$\phi$ directly from the azimuthal asymmetry between the scattered protons.
Forward proton tagging will also give access to a rich QCD program on hard diffraction
at high luminosities and to precision studies of $\gamma p$
and $\gamma \gamma$ interactions at center-of-mass energies never reached before.

\end{document}